\documentclass{osa-article}
\journal{osajournal}
\articletype{Research Article}

\usepackage{lineno}
\usepackage{amsmath}
\usepackage{newtxtext,newtxmath}
\usepackage{graphicx}
\usepackage{subfigure}
\usepackage{mathrsfs}
\usepackage{epsfig}
\newcommand{\ot}{\otimes}
\newcommand{\ket}[1]{|{#1}\rangle}
\newcommand{\bra}[1]{\langle{#1}|}
\newcommand{\bkt}[2]{\langle{#1}|{#2}\rangle}
\begin{document}

\title{The systematic error and the anomaly in the sensitivity and the probability of post-selection raised from the plural weak value}

\author{Jing-Hui Huang\authormark{1,5}, Xue-Ying Duan\authormark{2,3,4},  Guang-Jun Wang\authormark{2,3,4} and Xiang-Yun Hu\authormark{1,5,*}}

\address{\authormark{1} Institute of Geophysics and Geomatics, China University of Geosciences, Wuhan 430074, China\\
\authormark{2}School of Automation, China University of Geosciences, Wuhan 430074, China\\
\authormark{3}Hubei Key Laboratory of Advanced Control and Intelligent Automation for Complex Systems, Wuhan 430074, China\\
\authormark{4} Engineering Research Center of Intelligent Technology for Geo-Exploration, Ministry of Education, China\\
\authormark{5} State Key Laboratory of Geological Processes and Mineral Resources, China University of Geosciences, Wuhan, Hubei, 430074, China
}
\email{\authormark{*}xyhu@cug.edu.cn}

\begin{abstract}
Recently, weak value $\rm A_w$ derived in the pre- and post-selected weak measurement has been shown to be powerful in measuring minute physical effects. In principle, the decrease in the post-selection probability will increase the sensitivity. Besides, the sensitivity which characterizes the pointer position shift is proportional to the real part of $\rm A_w$, and the sensitivity which characterizes the pointer shift in momentum space is proportional to the imaginary part of $\rm A_w$. However, we find that the relationship between the post-selection probability and the sensitivity is true when $\rm A_w$ is a purely real number or a purely imaginary number. The plural $\rm A_w$ will lead to the abnormal behavior where the decrease in the post-selection probability reduce the sensitivity at certain pre- and post-selection. In addition, considering the device imperfections and the environmental instability in the WVA protocol, this anomaly raised from plural $\rm A_w$ will reduce the sensitivity and generate a systematic error of the measurement compared with the original scheme. 
Finally, three feasible methods are proposed to reduce these negative effects when the weak measurement inevitably changes into measurement with a plural weak value.
\end{abstract}

\section{Introduction}
High precision phase measurements with optical interferometers play a significant role in both classical measurements\cite{Li:08,Park:07} and quantum measurements\cite{2013Weak,doi:10.1126/science.1202218,PhysRevLett.124.173602,Dooley:15}. Numerous studies show that the quantum measurements with weak value amplification(WVA) can beat the classical schemes with high sensitivity and good robustness\cite{2009Ultrasensitive,PhysRevA.102.042601,PhysRevLett.126.020502,Huang2021}. The concept of weak value in weak measurement was first derived from the work on the arrow of time in quantum theory by  Aharonov, Albert and Vaidman in 1988\cite{AAV}. In particular, the weak value $\rm A_w$ derived by a weak measurement on pre-selected and post-selected can be arbitrarily large, where the coupling strength between the system and the meter is sufficiently small\cite{PhysRevLett.105.010405}. Thus, the famous example of the weak measurement of a spin particle leading to a value of 100\cite{AAV} has opened up a gate for the WVA in the estimation of small parameters, such as ultra-sensitive beam deflection measurement\cite{2009Ultrasensitive}, velocity measurements\cite{2013Weak}, ultra-small polarization rotation\cite{DELIMABERNARDO20142029}, rotation velocity\cite{Huang2021}, single-photon nonlinearity\cite{PhysRevLett.107.133603}, precision thermometry\cite{PhysRevA.102.012204}.

In principle, the information is gained at principle at the price of disturbing the system through the interaction and the system will be projected onto certain eigenstates in quantum mechanics. However, in weak measurement, the disturbance to the system is reduced at the cost of a similar reduction in the amount of information provided by the measurement\cite{Hallaji2017}. In this case, the probe shifts, averaged over many measurement repetitions, have been amplified by the weak value $\rm A_w$ of the observable $\hat{A}$: $A_{w}:={\bra{\Phi_{f}}\hat{A}\ket{\Phi_{i}}}/{\bkt{\Phi_{f}}{\Phi_{i}}},$ where $\ket{\Phi_{i}}$ and $\ket{\Phi_{f}}$ are the pre-selected and post-selected states of the system. It has been suggested that a standard interferometer greatly outperforms weak measurement in a scenario involving a purely real weak value, but the measurement of a phase using a purely imaginary weak value can outperform a standard interferometer\cite{PhysRevLett.105.010405}. Hence, anomalous works\cite{10.1109/JPHOT.2019.2942718,PhysRevA.102.023701,Yin2021,PhysRevA.103.032212} of amplifying the detector signals based on weak measurement with a purely imaginary weak value. So far, the weak measurements involving a plural weak value have been seldom reported.

In realistic optical interferometers, the pre-selected and post-selected states of the system are normally obtained by polarizers or the combination of wave-plate and polarizer. Due to device imperfections and environmental instability, the pre-selected state $\ket{\Phi_{i}}$ and post-selected state $\ket{\Phi_{f}}$ of the system may deviate from the desired quantum states. In our work, we find that this deviation will change the purely real or imaginary weak value to a plural one. Specifically, we study three measurements for detecting a small phase shift introduced by a birefringent element involving (a) weak measurements with a purely real $\rm A_w$, (b) weak measurements with a purely imaginary $\rm A_w$, (c) weak measurements with a plural $\rm A_w$. As a result, in the scheme (a), the lower probability of post-selection $\mathcal{P}=|{\bkt{\Phi_{f}}{\Phi_{i}}}|^{2}$ increases the sensitivity(proportional to the real part $\rm Re$ $\rm [A_w]$ of $\rm A_w$); and the lower $\mathcal{P}$ leads to the higher sensitivity(proportional to the imaginary part $\rm Im$ $\rm [A_w]$ of $\rm A_w$) in the scheme (b). Nevertheless, an anomaly in the sensitivity and the probability of post-selection will be raised from the plural weak value, and the anomaly will cause systematic errors when $\mathcal{P}$ is small enough.

This paper is organized as follows. In section 2, we propose the theoretical scheme of (a) a weak measurement with a purely real $\rm A_w$, (b) a weak measurement with a purely imaginary $\rm A_w$ and (c) a weak measurement a plural $\rm A_w$ to measure time delay induced by a birefringent element. Then in section 3, numerically, we show these results of the three schemes and analyze the anomaly in the sensitivity and the probability of post-selection in scheme 3. Then, in section 4, we analyze the systematic error caused by the anomaly and put forward some possible solutions to reduce the error. Finally, a brief conclusion is presented in section 5.

\section{Schemes of weak measurements involving weak values with different complex properties }

\begin{figure*}[htp!]
	\centering
	%\vspace{-0.2cm}
\subfigure
{
	%\begin{minipage}{0.5\linewidth}
	\vspace{-0.2cm}
	\begin{minipage}{6.5cm}
	\centering
	\centerline{\includegraphics[scale=0.35,angle=0]{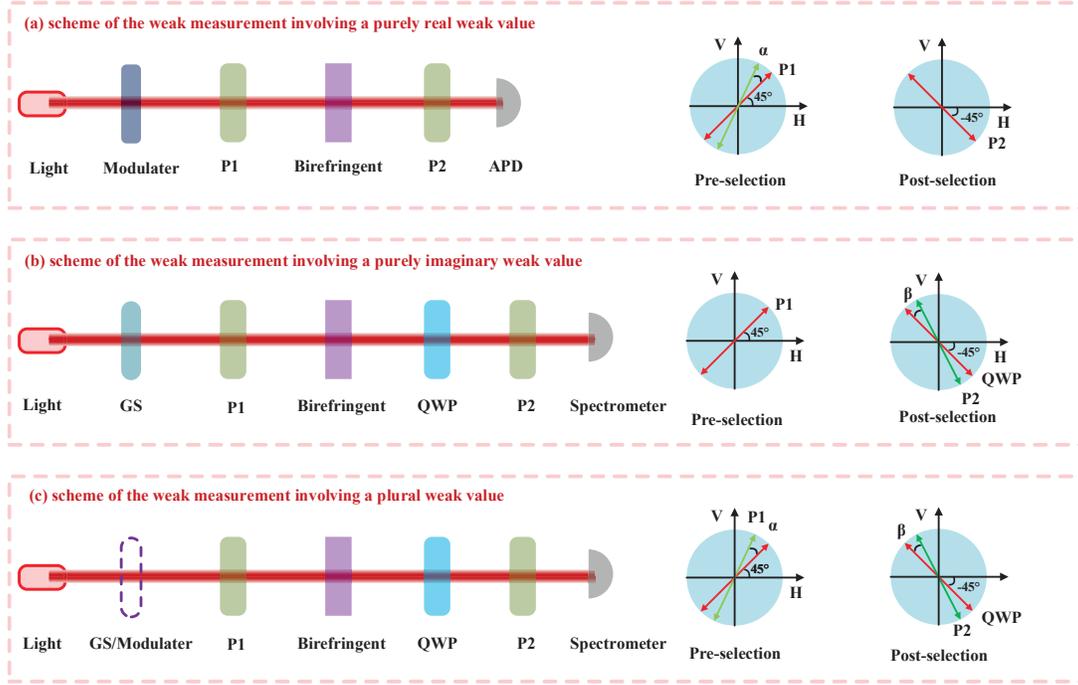}}
	\end{minipage}
}
\vspace*{0mm} \caption{\label{Schemes_model}Schemes of weak measurements involving weak values with different complex properties.  }
\end{figure*}
To compare the three schemes of weak measurements involving a purely real $\rm A_w$, a purely imaginary $\rm A_w$ and a plural $\rm A_w$, we propose three schemes of a two-level system($\ket{\Phi}$ is a component of the horizontal polarized state $\ket{H}$ and the vertical polarized state $\ket{V}$) in Fig. \ref{Schemes_model} to detect the same time delay $\tau$ induced by a birefringent element with different pre- and post-selected states of the system, where the time delay $\tau$ can be measured serves as the coupling strength of the interaction between the system $\ket{\Phi}$ and the probe $\ket{\Psi}$ in standard weak measurements. Thus, the interaction Hamiltonian $\hat{H}=\tau \hat{A}\ot \hat{p}$ leads to a unitary operator
\begin{eqnarray}
\label{interaction-Hamiltonian}
\hat{U}=e^{-i \int \hat{H} dt}=e^{-i\tau \hat{A}\ot \hat{p}} \, ,
\end{eqnarray}
where $\hat{A}=\ket{H} \bra{H}-\ket{V} \bra{V} $ is the observable operator, and $\hat{p}$ is the probe momentum operator conjugate to the position operator $\hat{q}$. In principle, the time delay $\tau$ can be obtained by the shifts of the probe function $\ket{\Psi}$, and the probe function after the post-selection will take the form:
\begin{eqnarray}
\label{inter_prob_final}
\ket{\Psi_{f}} &=&\bra{\Phi_{f}} \hat{U}\ket{\Phi_{i}}\ket{\Psi_{i}} \nonumber \\
&=&\bra{\Phi_{f}}e^{-i\tau\hat{A}\ot \hat{p}}\ket{\Phi_{i}}\ket{\Psi_{i}} \nonumber \\
&=&\bra{\Phi_{f}}\left[ 1-i\tau\hat{A}\ot \hat{p}\right]\ket{\Phi_{i}}\ket{\Psi_{i}}+O(\tau ^{2}) \nonumber \\
&=&\bkt{\Phi_{f}}{\Phi_{i}}\left[ 1-i\tau A_{w}\hat{p}\right]\ket{\Psi_{i}}+O(\tau ^{2}) \nonumber \\
&=&\bkt{\Phi_{f}}{\Phi_{i}}e^{-i\tau A_{w}\hat{p}}\ket{\Psi_{i}}+O(\tau ^{2})\, ,
\end{eqnarray}
where $O(\tau ^{2})$ is normally neglected when the coupling strength is weak for $\tau|A_{w}|\ll 1$. Choosing the Gaussian form with the variance $W^{-2}$ for the initial probe function $\ket{\Psi_{i}}$ in the position or momentum spaces, the weak value can be obtained from the shifts of the expectation value of the probe momentum p and the probe position q as \cite{PhysRevA.85.052110} 
\begin{equation}
\label{delta_q}
\Delta\langle\hat{q}\rangle=\frac{\int d q q\left|\left\langle q \mid \Psi_{f}\right\rangle\right|^{2}}{\int d q\left|\left\langle q \mid \Psi_{f}\right\rangle\right|^{2}}=\tau {\rm Re} \,\rm [A_{w}] \,,
\end{equation} 
\begin{equation}
\label{delta_p}
\Delta\langle\hat{p}\rangle=\frac{\int d p p\left|\left\langle p \mid \Psi_{f}\right\rangle\right|^{2}}{\int d p\left|\left\langle p \mid \Psi_{f}\right\rangle\right|^{2}}=2\tau W^{2} {\rm Im} \, \rm [A_{w}] \, .
\end{equation} 
Equation (\ref{delta_q}) shows that the shifts of the expectation value of the probe position are proportional to the real part $\rm  Re \, [A_w]$ of the weak value. And equation (\ref{delta_p}) shows that the shifts of the expectation value of the probe momentum are proportional to the imaginary part $\rm  Im \, [A_w]$ of the weak value. As a trade-off, the probability $\mathcal{P}=|{\bkt{\Phi_{f}}{\Phi_{i}}}|^{2}$ of obtaining a post-selected state that is almost orthogonal to the pre-selected state is very small. Note that the complex properties of $\rm   \, [A_w]$ are defined by the pre- and post-selected states of the system. In the following, we will show the theoretical scheme of (a) a weak measurement with a purely real $\rm A_w$, (b) a weak measurement with a purely imaginary $\rm A_w$ and (c) a weak measurement a plural $\rm A_w$ to measure time delay induced by a birefringent element.

\subsection{Scheme of weak measurements involving a purely real weak value  }
In order to realize the weak measurement involving a purely real $\rm A_w$, the experimental model is shown in Fig. \ref{Schemes_model}(a). Optical modulator generates a Gaussian-shaped pulse in the time domain, which travels through the first polarizer (P1) that pre-selected the system in the polarized state $\ket{\Phi_{i}}= {\rm sin} (\frac{\pi}{4}+\alpha) \ket{H}+ {\rm cos} (\frac{\pi}{4}+\alpha)\ket{V}$, after which it enters a birefringent element to perturb the system weakly. Afterwards, the second polarizer (P2) post-select the system at $\ket{\Phi_{f}}={\rm sin} (-\frac{\pi}{4}) \ket{H}+ {\rm cos} (-\frac{\pi}{4})\ket{V}$. The corresponding time shifts in the peak out intensity $\delta t=\tau {\rm Re} \,\rm [A_{w}]$ are detected by an avalanche photodiode (APD), which is proportional to $ {\rm Re} \,\rm [A_{w}]$. Note that the $ \rm [A_{w}]$ can be calculated by
\begin{equation}
\label{weak_value_1}
A_{w,1}=\frac{\bra{\Phi_{f}}\hat{A}\ket{\Phi_{i}}}{\bkt{\Phi_{f}}{\Phi_{i}}}=\frac{-{\rm sin} (\frac{\pi}{4}+\alpha) - {\rm cos} (\frac{\pi}{4}+\alpha)}{-{\rm sin} (\frac{\pi}{4}+\alpha) + {\rm cos} (\frac{\pi}{4}+\alpha)} ={\rm cot} \alpha \, ,
\end{equation} 
and Eq. (\ref{weak_value_1}) indicates $ {\rm Re} \,\rm [A_{w}]$ = $ {\rm Ab} \,\rm [A_{w}]$, where $ {\rm Ab} \,\rm [A_{w}]$ represents the modulus of $\rm A_w$. Meanwhile, the  probability of post-selection is given by
\begin{equation}
\label{post_selected_1}
|{\bkt{\Phi_{f}}{\Phi_{i}}}|^{2}_{1}=  {\rm sin}^{2} \alpha \, .
\end{equation} 
In scheme (a), the sensitivity $\delta t / \tau $ is proportional to $\rm Re \,\rm [A_{w}]$, and the relationship between $\rm A_w$ and the probability of post-selection is shown in Fig. \ref{result_numberical}(a).

\subsection{Scheme of weak measurements involving a purely imaginary weak value }
\begin{figure*}[htp!]
	\centering
	%\vspace{-0.2cm}
\subfigure
{
	%\begin{minipage}{0.5\linewidth}
	\vspace{-0.2cm}
	\begin{minipage}{6.5cm}
	\centering
	\centerline{\includegraphics[scale=0.55,angle=0]{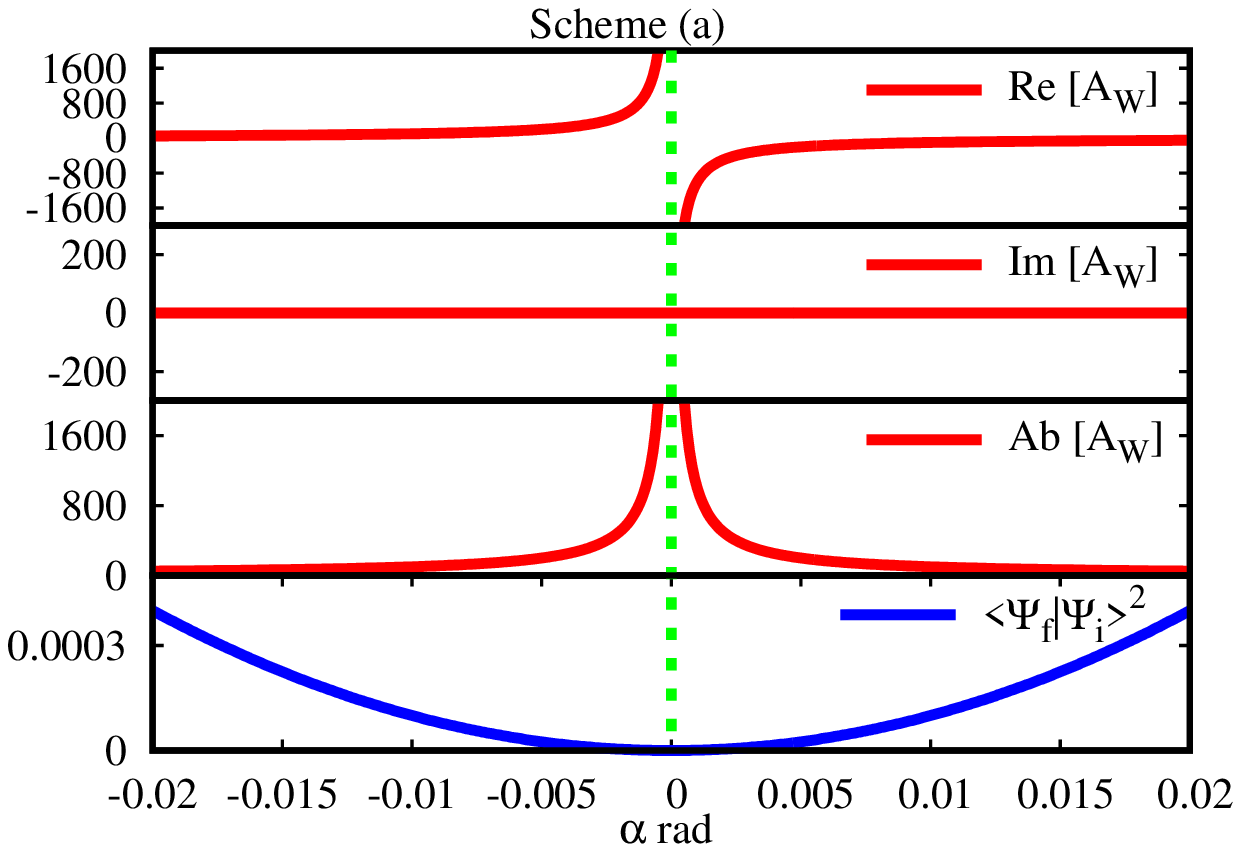}}
	\end{minipage}
}
\vspace{-0.2cm}
\subfigure
{
	%\begin{minipage}{0.5\linewidth}
	\begin{minipage}{6.4cm}
	\centering
	\centerline{\includegraphics[scale=0.55,angle=0]{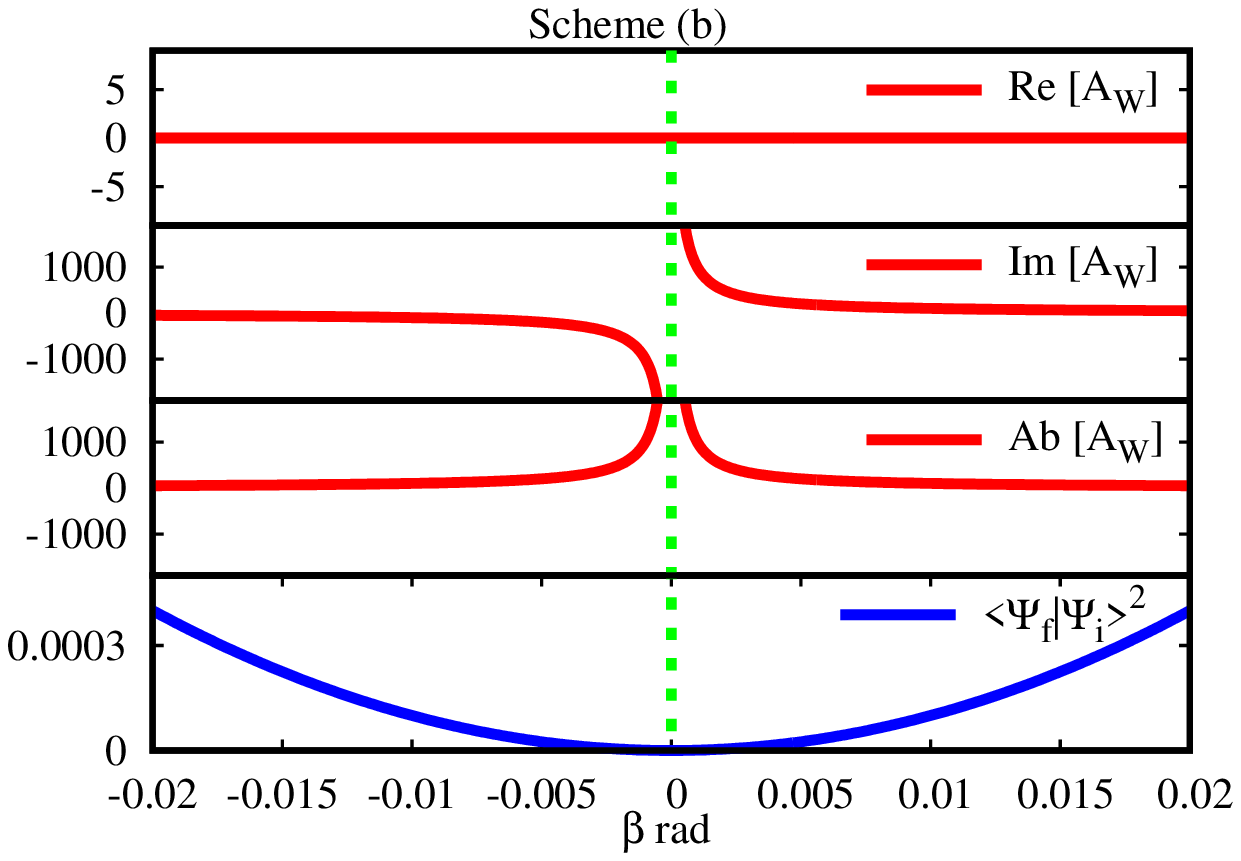}}
	\end{minipage}
}
\vspace{-0.2cm}

\subfigure
{
	%\begin{minipage}{0.5\linewidth}
	\begin{minipage}{6.5cm}
	\centering
	\centerline{\includegraphics[scale=0.55,angle=0]{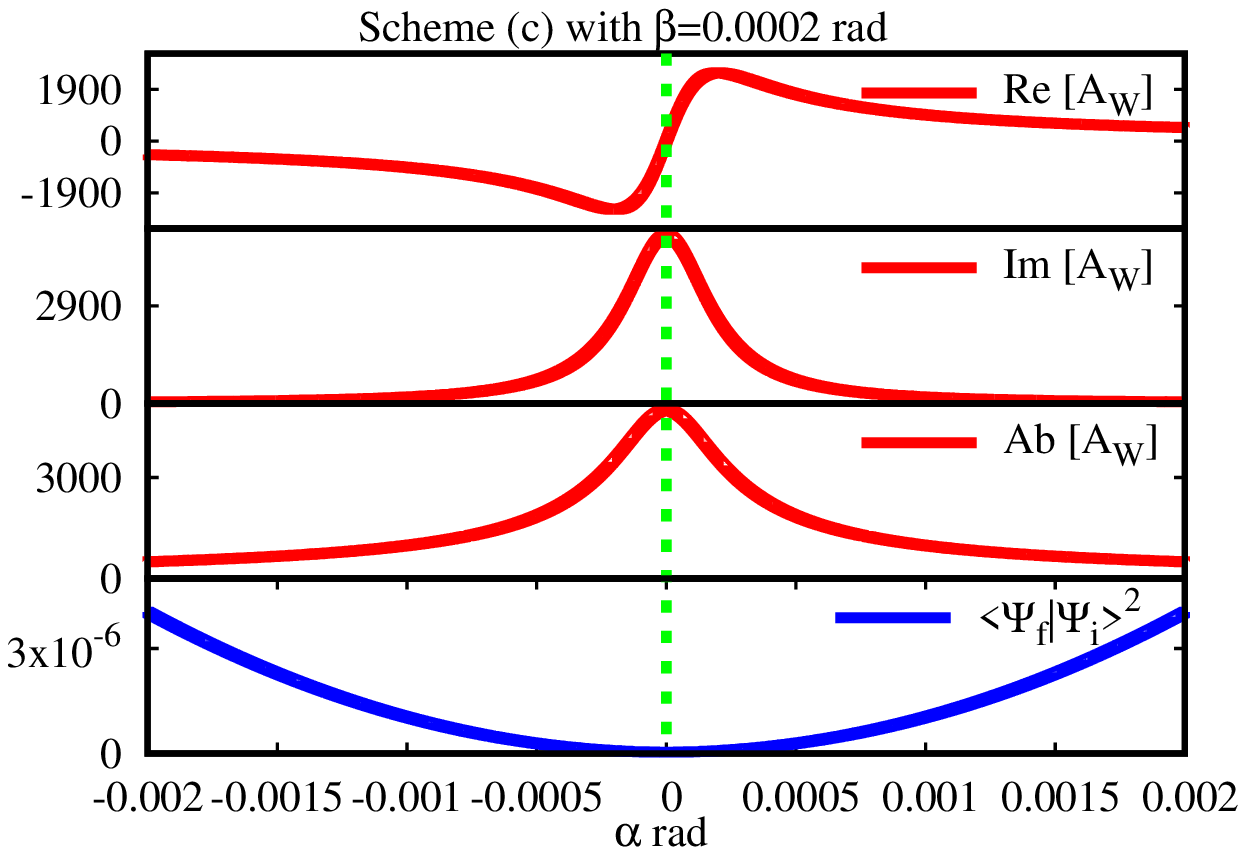}}
	\end{minipage}
}
\subfigure
{
	%\begin{minipage}{0.5\linewidth}
	\begin{minipage}{6.4cm}
	\centering
	\centerline{\includegraphics[scale=0.55,angle=0]{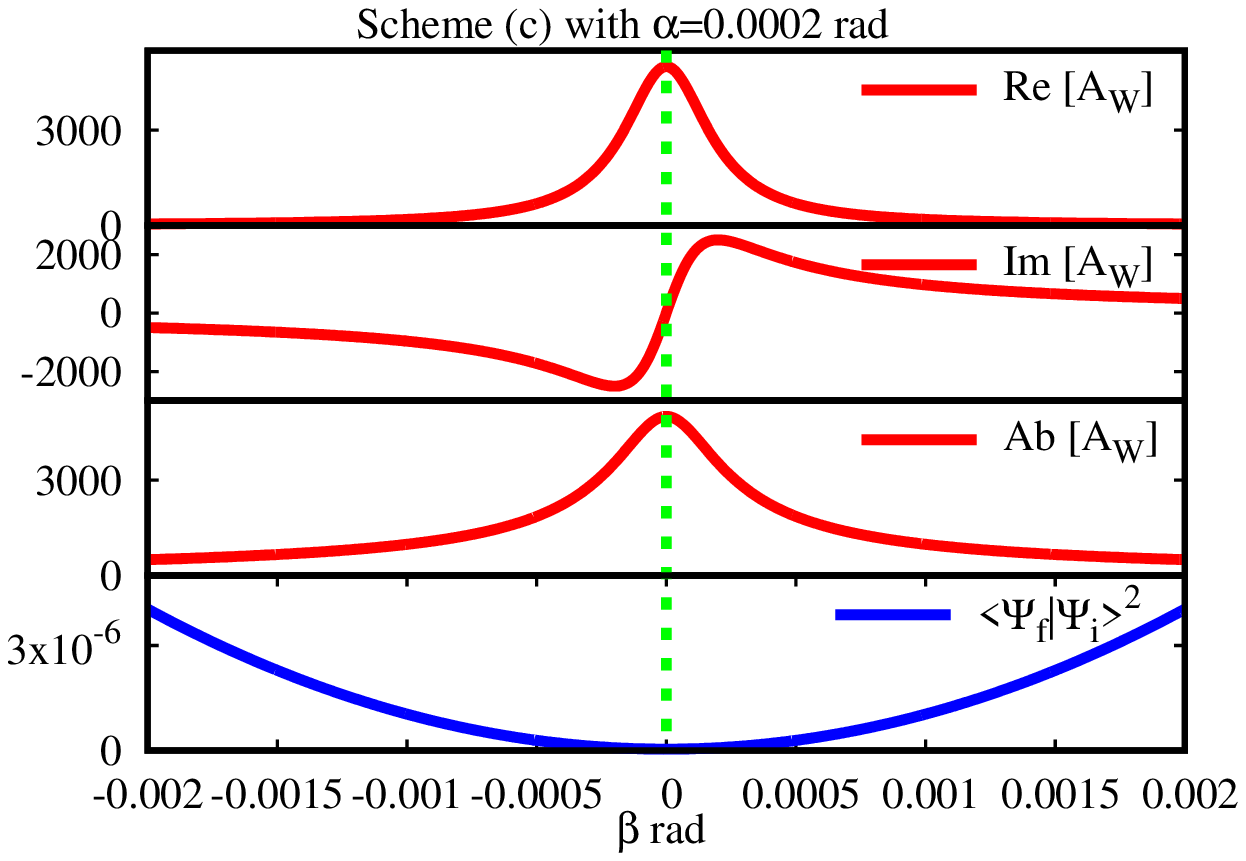}}
	\end{minipage}
}
\vspace{-0.2cm}

\subfigure
{
	%\begin{minipage}{0.5\linewidth}
	\begin{minipage}{6.5cm}
	\centering
	\centerline{\includegraphics[scale=0.55,angle=0]{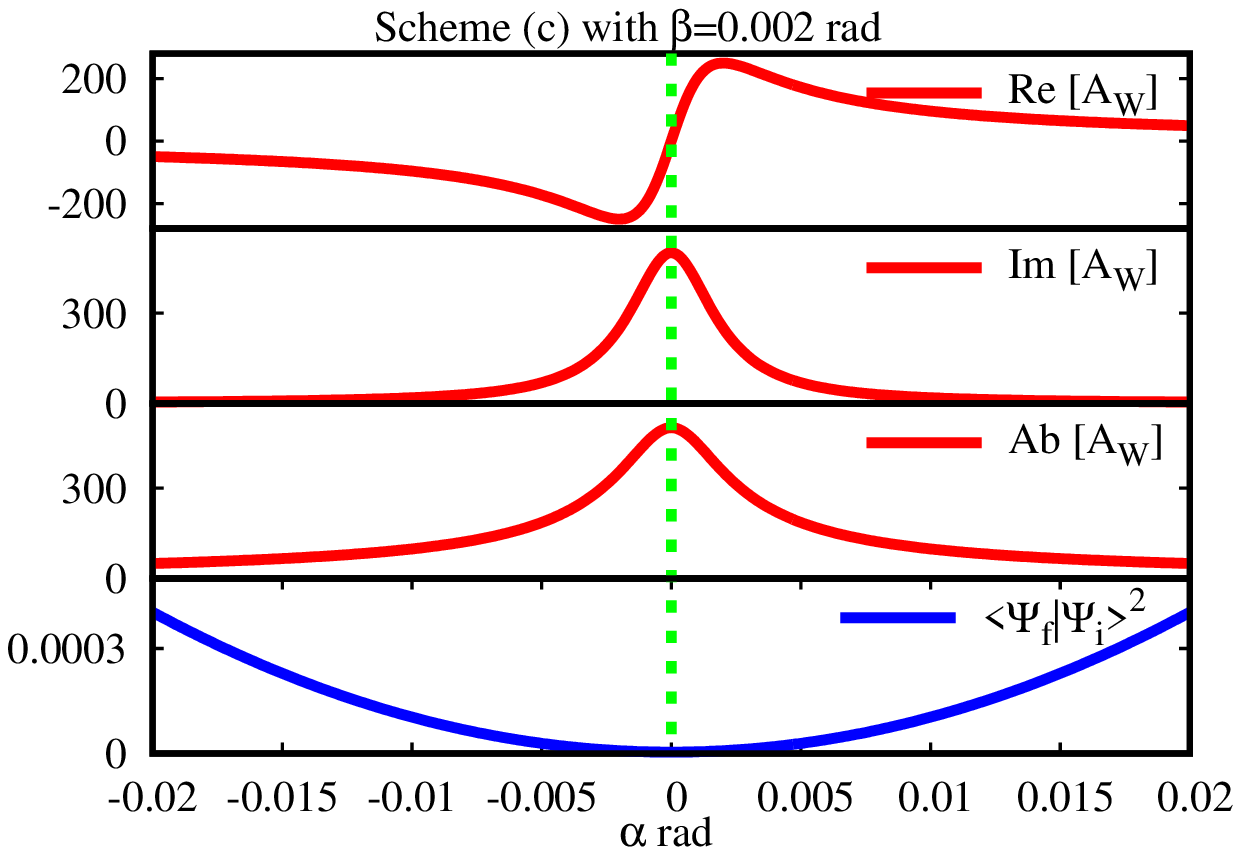}}
	\end{minipage}
}
\subfigure
{
	%\begin{minipage}{0.5\linewidth}
	\begin{minipage}{6.4cm}
	\centering
	\centerline{\includegraphics[scale=0.55,angle=0]{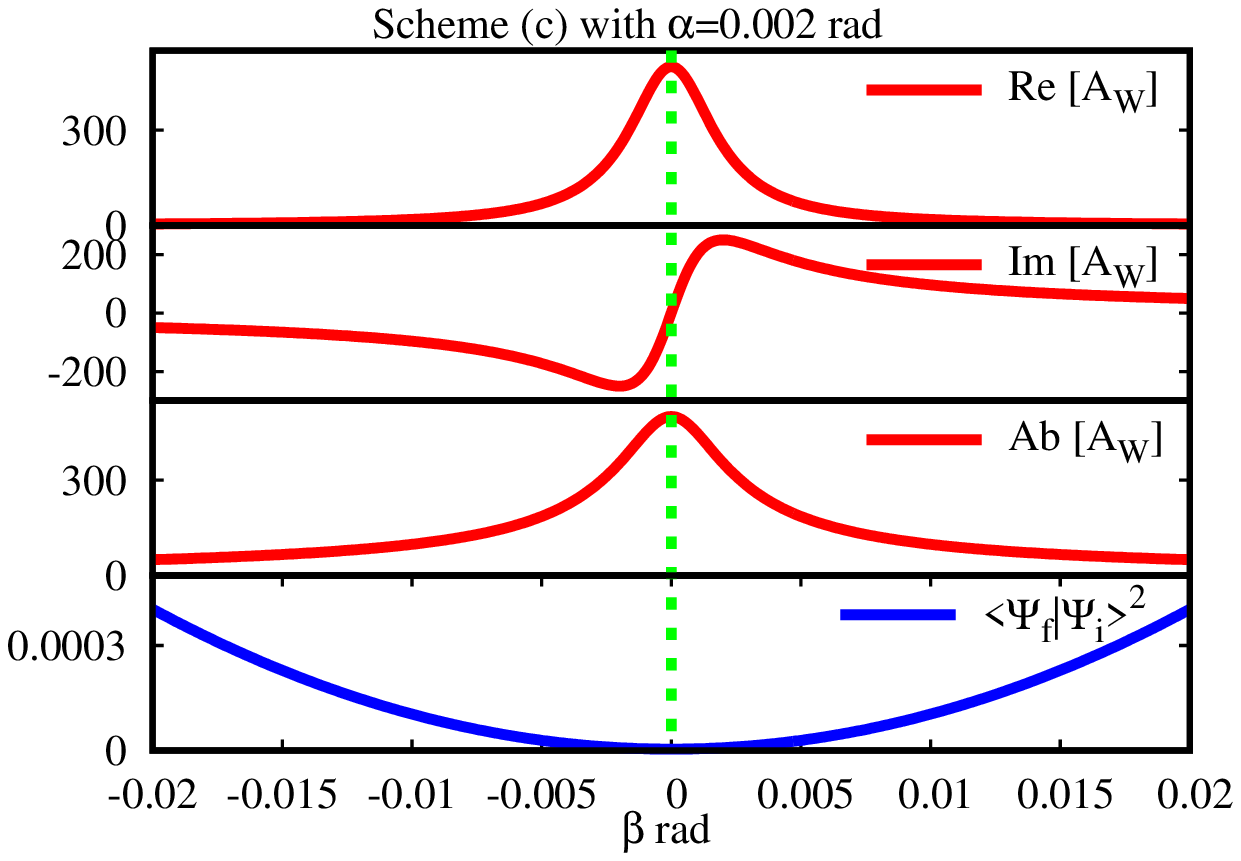}}
	\end{minipage}
}
\vspace{-0.2cm}

\vspace*{0mm} \caption{\label{result_numberical}The results of our simulation experiment with different models.  }
\end{figure*}

Fig. \ref{Schemes_model}(b) depicts the scheme involving a purely imaginary $\rm A_w$ in momentum space. Normally, a white laser and a following Gaussian filter together constitute the probe state. Then, the light beam passes the first polarizer (P1) which pre-selected the system at the state $\ket{\Phi_{i}}= \frac{1}{\sqrt{2}} (\ket{H}+ \ket{V})$.  perturb the system weakly.  A birefringent element introduces a time delay $\tau$, which corresponds to the coupling strength. Following these works\cite{doi:10.1063/1.4976312,PhysRevA.102.023701,PhysRevA.103.032212} to realize the post-selection, a quarter-wave plate (QWP) and the second polarizer (P2) post-select the system at the state,
\begin{equation}
\ket{\Phi_{f}} \simeq \frac{1}{\sqrt{2}}\left(\begin{array}{cc}
1 & -i \\
-i & 1
\end{array}\right)\left(\begin{array}{c}
\cos (\beta-\pi / 4) \\
\sin (\beta-\pi / 4)
\end{array}\right) =\frac{1}{\sqrt{2}}[\exp (-i \beta)|H\rangle-\exp (i \beta)|V\rangle] \,,
\end{equation}
where $\beta $ is the angle between the optical axes of P2 and QWP. Hence, $\rm A_w$ and the probability of post-selection can be obtained immediately by that:
\begin{equation}
\label{weak_value_2}
A_{w,2}=\frac{\bra{\Phi_{f}}\hat{A}\ket{\Phi_{i}}}{\bkt{\Phi_{f}}{\Phi_{i}}}=-i {\rm cot} \beta \, ,
\end{equation} 
\begin{equation}
\label{post_selected_2}
|{\bkt{\Phi_{f}}{\Phi_{i}}}|^{2}_{2}=  {\rm sin}^{2} \beta \, .
\end{equation} 
Note that scheme (2) involves a purely imaginary weak value with a probe in the frequency(momentum) domain. Therefore, the sensitivity $\delta p / \tau $ is proportional to $\rm Im \,\rm [A_{w}]={\rm cot} \beta$, and we display the relationship between $\rm A_w$ and the probability of post-selection in Fig. \ref{result_numberical}(b).

\subsection{Scheme of weak measurements involving  a plural weak value  }
In this subsection, the scheme of a weak measurement a plural $\rm A_w$ is shown in Fig. \ref{Schemes_model}(c). The main difference comparing to scheme (2) is the pre-selected state of the system controlled by P1:
\begin{equation}
\label{pre_selected_3}
\ket{\Phi_{i}}=  {\rm sin}(\alpha+\frac{\pi}{4}) \ket{H}+ {\rm cos}(\alpha+\frac{\pi}{4}) \ket{V}\, .
\end{equation} 
and the pre-selected sate $\ket{\Phi_{f}}=\frac{1}{\sqrt{2}}[\exp (-i \beta)|H\rangle-\exp (i \beta)|V\rangle]$ of the system keeps the same with the form in scheme (2). Due to the deviation on the pre-selected state of the system, $\rm A_w$ and $|{\bkt{\Phi_{f}}{\Phi_{i}}}|^{2}$ are calculated as follows:
\begin{equation}
\label{weak_value_3}
A_{w,3}=\frac{\bra{\Phi_{f}}\hat{A}\ket{\Phi_{i}}}{\bkt{\Phi_{f}}{\Phi_{i}}}=\frac{({\rm sin}\alpha+{\rm cos}\alpha){\rm exp}(-i\beta)+(-{\rm sin}\alpha+{\rm cos}\alpha){\rm exp}(i\beta)}
{({\rm sin}\alpha+{\rm cos}\alpha){\rm exp}(-i\beta)-(-{\rm sin}\alpha+{\rm cos}\alpha){\rm exp}(i\beta)} \, ,
\end{equation} 
\begin{equation}
\label{post_selected_3}
|{\bkt{\Phi_{f}}{\Phi_{i}}}|^{2}_{3}=\frac{1}{4} [
({\rm sin}\alpha+{\rm cos}\alpha)^{2}+
(-{\rm sin}\alpha+{\rm cos}\alpha)^{2}
-2({\rm cos}^{2}\alpha-{\rm sin}^{2}\alpha) {\rm cos}2\beta
] \, .
\end{equation} 
The theoretical formula (\ref{weak_value_3}) indicates that the deviation on $\ket{\Phi_{f}}$ will lead to a plural weak value in the scheme (3). 
we consider a deviation on the pre-selected state of the system through introducing a small deflection $\alpha$ of the first polarizer (P1), and a deviation on the post-selected state of the system through introducing a small deflection $\beta$ of the second polarizer (P2).
The scheme with the light followed by optical modulator is compared to the scheme (1) involving a temporal probe. Note that when $\beta \to 0$, the scheme (3) will be the same as the scheme (1) and lead to the same $A_w$ and $|{\bkt{\Phi_{f}}{\Phi_{i}}}|^{2}$:
\begin{equation}
\left\{\begin{array}{l}
A_{w, 3}=A_{w, 1} \\
|{\bkt{\Phi_{f}}{\Phi_{i}}}|^{2}_{3}=|{\bkt{\Phi_{f}}{\Phi_{i}}}|^{2}_{1}
\end{array} \quad \text { when } \beta \rightarrow 0 \, .\right.
\end{equation}

In addition, another scheme with the light followed by a Gaussian filter is compared to the scheme (2) involving a frequency-domain probe. Accordingly, when $\beta \to 0$, the scheme (3) will be the same as the scheme (2) and lead to the same $A_w$ and $|{\bkt{\Phi_{f}}{\Phi_{i}}}|^{2}$:
\begin{equation}
\left\{\begin{array}{l}
A_{w, 3}=A_{w, 2} \\
|{\bkt{\Phi_{f}}{\Phi_{i}}}|^{2}_{3}=|{\bkt{\Phi_{f}}{\Phi_{i}}}|^{2}_{2}
\end{array} \quad \text { when } \alpha \rightarrow 0 \, .\right.
\end{equation}

On the other hand, the scheme (a) will translate the scheme (c) mathematically due to the deviation. Although this translation can not occur in real optical systems as shown in Fig. \ref{Schemes_model}, in this paper we just compare it with the weak measurement with a purely imaginary $\rm A_w$. Nevertheless, the translation from the scheme (c) to the scheme (d) will occur in both mathematically and in real optical systems. In the next section, we will show how the deviation of the states of the system affects the weak measurement. 

\section{Results}
In order to discuss the impact of the deviation on the weak measurement in position or momentum space. We numerically simulate the results of scheme(a), scheme(b) and scheme(c). Thereinto, the scheme(c) with the deflection ($\beta$=0.0002 rad and 0.002 rad) of P2 serves as the deviation of scheme(a) involving a weak measurement with a purely real $\rm A_w$, and the scheme(c) with the deflection ($\alpha$=0.0002 rad and 0.002 rad) of P1 serves as the deviation of the scheme(b) involving a weak measurement with a purely imaginary $\rm A_w$. The relevant results are shown in Fig. \ref{result_numberical}.

Comparing the results(left panel of Fig. \ref{result_numberical}) of the scheme(c) with the deflection ($\beta$=0.0002, 0.002 rad) and the scheme(a) with a purely real $\rm A_w$, we find that the deflection on $\beta$ will lead to the weak measurement involving a plural weak value. At this time the function of $\rm Re$ $\rm [A_w]$ in the scheme (c) is not a monotone function of $\alpha$, and the anomaly in the sensitivity and the probability of post-selection is raised when $\alpha$ is small on the order of the deflection. For instance, in the scheme (c) with $\beta$= 0.0002 rad, the sensitivity (proportional to $\rm Re$ $\rm [A_w]$) decreases as the probability of post-selection $\mathcal{P}=|{\bkt{\Phi_{f}}{\Phi_{i}}}|^{2}$ decreases when $\alpha$ < 0.0002 rad, while $\rm Re$ $\rm [A_w]$ increases as  $\mathcal{P}$ decreases when $\alpha$ < 0.0002 rad. Although the anomaly in scheme (c) with $\beta$= 0.0002 rad will not appear due to the value of $\alpha$ is going to be much larger than 0.0002 rad in a real experiment, the anomaly in scheme (c) with $\beta$= 0.002 rad or bigger $\beta$ will occur when $\alpha$ is set on the order of the deflection($\beta$= 0.002 rad).  

\begin{figure*}[htp!]
	\centering
	%\vspace{-0.2cm}
\subfigure
{
	%\begin{minipage}{0.5\linewidth}
	\vspace{-0.2cm}
	\begin{minipage}{6.5cm}
	\centering
	\centerline{\includegraphics[scale=0.55,angle=0]{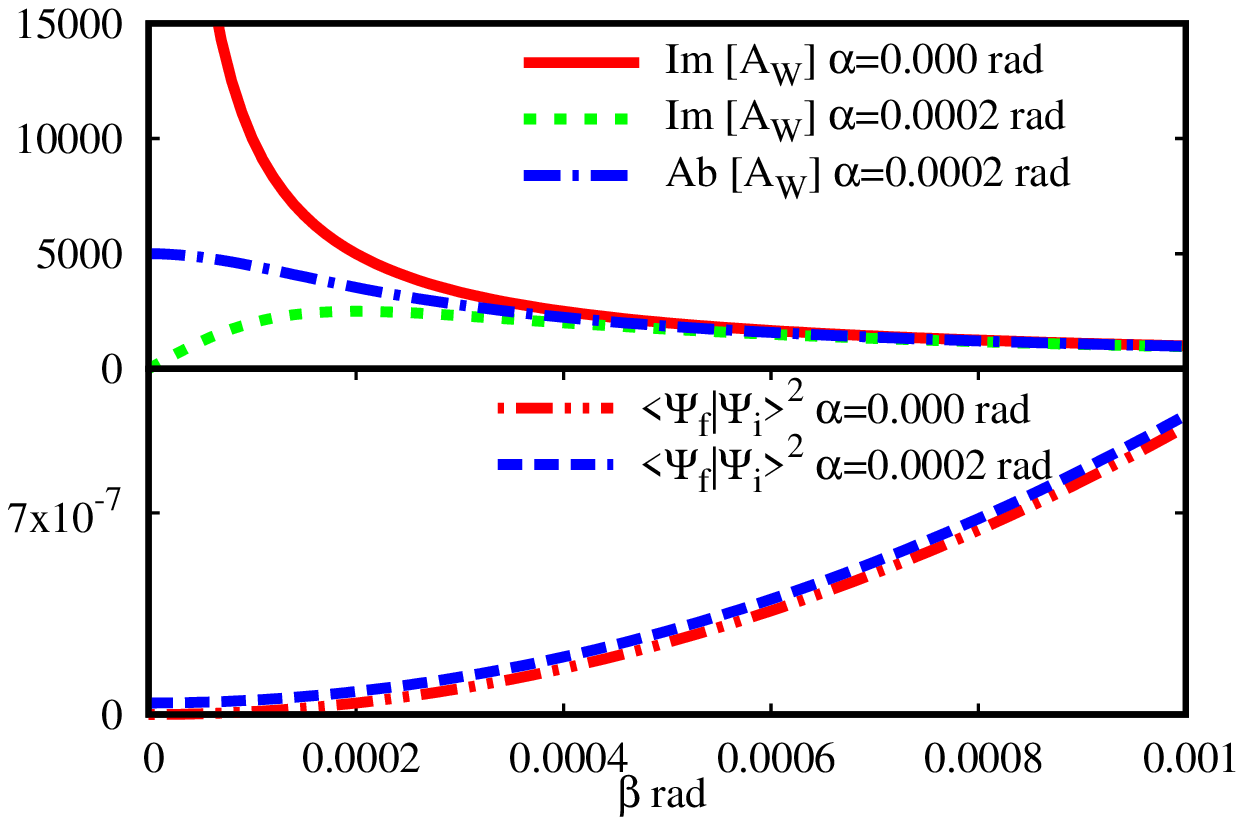}}
	\end{minipage}
}
\vspace{-0.2cm}
\subfigure
{
	%\begin{minipage}{0.5\linewidth}
	\begin{minipage}{6.4cm}
	\centering
	\centerline{\includegraphics[scale=0.55,angle=0]{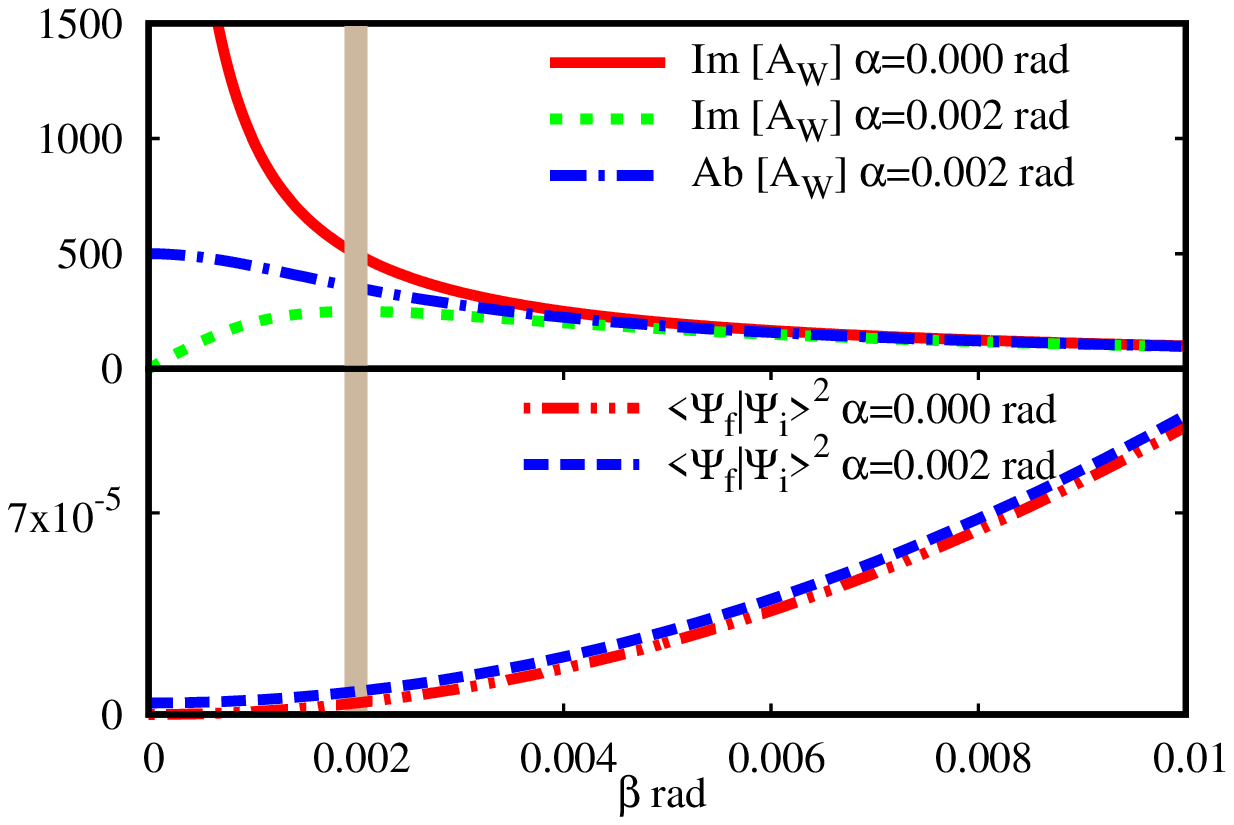}}
	\end{minipage}
}
\vspace{-0.2cm}
\vspace*{0mm} \caption{\label{result_numberical_compare}The comparison of the scheme(c) with the deflection ($\alpha$=0.0002, 0.002 rad) and the scheme(b) with a purely imaginary $\rm A_w$, where $\alpha$=0.000 rad is corresponding to the scheme (b). }
\end{figure*}

Meanwhile, similar results of the comparison of the scheme(c) with the deflection ($\alpha$=0.0002, 0.002 rad) and the scheme(b) with a purely imaginary $\rm A_w$ can obtain from Fig. \ref{result_numberical}. In particular, the contrast of $\rm Im$ $\rm [A_w]$ and $\rm Ab$ $\rm [A_w]$with different $\alpha$ are shown in Fig. \ref{result_numberical_compare}, where $\rm Im$ $\rm [A_w]$ with $\alpha$= 0.000 rad is corresponding to the result of the scheme(b) with a purely imaginary $\rm A_w$. Considering $\beta$ set to be 0.002 rad in order to obtain the value of $\rm Im$ $\rm [A_w]$, there is no difference between $\rm Im$ $\rm [A_w]$ with $\alpha$= 0.000 rad and $\rm Im$ $\rm [A_w]$ with $\alpha$= 0.0002 rad. Nevertheless, the significant difference between $\rm Im$ $\rm [A_w]$ with $\alpha$= 0.000 rad and $\rm Im$ $\rm [A_w]$ with $\alpha$= 0.002 rad appears. Note that the deflection on P1 in the scheme (b) will happen in real experiments due to the device imperfections and environmental instability, and this deflection on P1 will change the scheme (b) into the scheme (c) involving a plural weak value. And this anomaly raised from the plural $\rm A_w$ will cause sensitivity decreases as compared with original design indicators.

In addition, an interesting conclusion is obtained from the results with a plural weak value. In Fig. \ref{result_numberical}, we find that when the weak measurement is realized involving a plural weak value, the probability of post-selection $\mathcal{P}$ is inversely proportional to the modulus of the weak value $\rm Ab$ $\rm [A_w]$, rather than $\rm Re$ $\rm [A_w]$ or $\rm Im$ $\rm [A_w]$. And this anomaly goes against previous experience which reducing the probability of post-selection $\mathcal{P}$ can improve the sensitivity of the weak measurement with a pure real $\rm [A_w]$ or a pure imaginal $\rm [A_w]$. We hope that this anomaly can attract the attention of the WVA community, from both experimental and theoretical interests. 

\section{Systematic error analysis}
\begin{figure*}[htp!]
	\centering
	%\vspace{-0.2cm}
\subfigure
{
	%\begin{minipage}{0.5\linewidth}
	\vspace{-0.2cm}
	\begin{minipage}{6.5cm}
	\centering
	\centerline{\includegraphics[scale=0.55,angle=0]{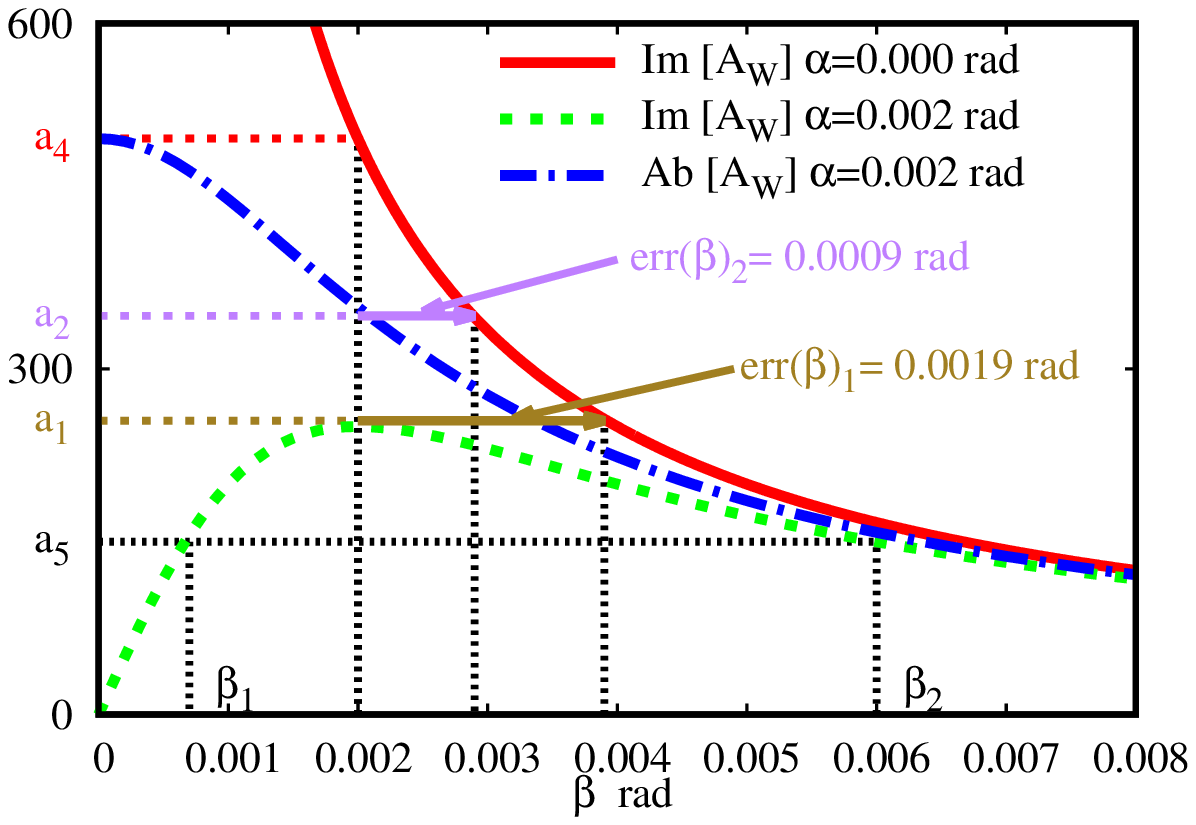}}
	\end{minipage}
}
\vspace{-0.2cm}
\subfigure
{
	%\begin{minipage}{0.5\linewidth}
	\begin{minipage}{6.4cm}
	\centering
	\centerline{\includegraphics[scale=0.55,angle=0]{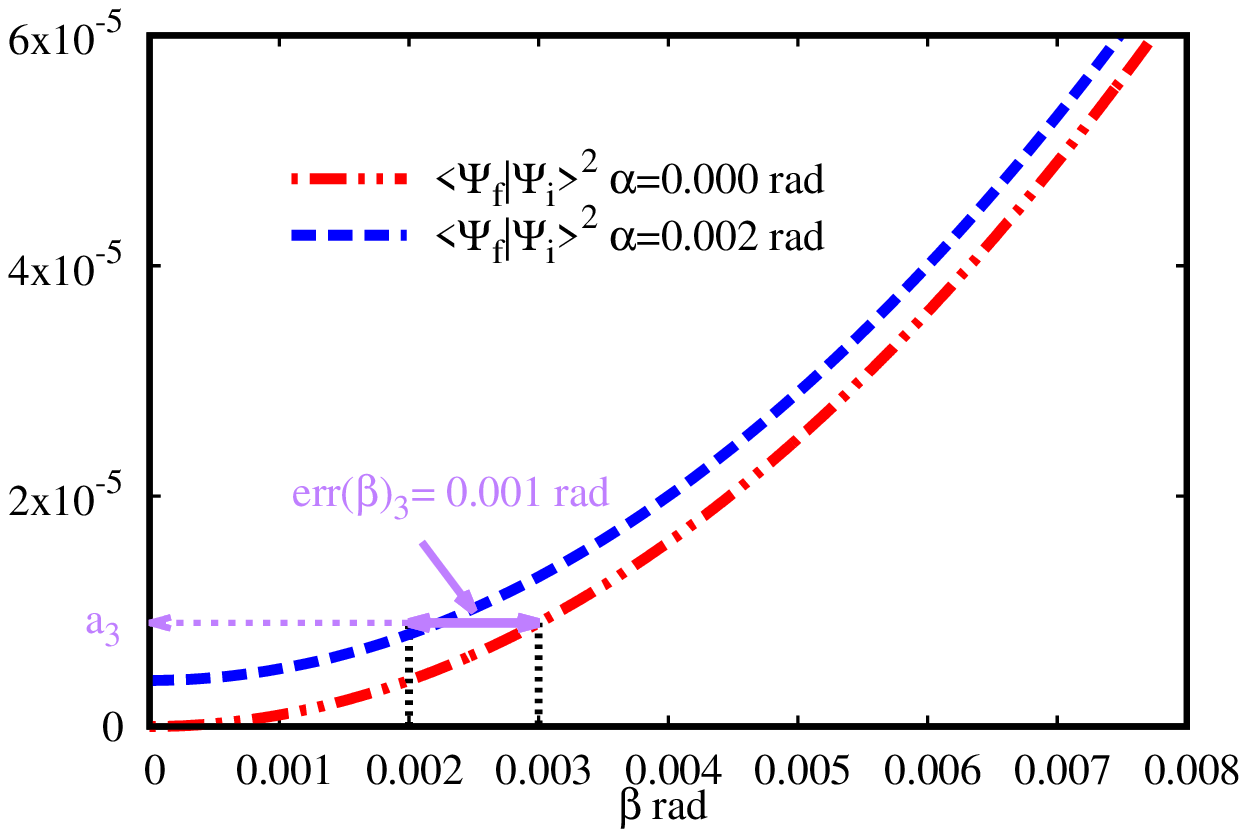}}
	\end{minipage}
}
\vspace{-0.2cm}
\vspace*{0mm} \caption{\label{Fig-system-error}Systematic error err($\beta$) on the measurement of $\beta$= 0.002 rad with the anomaly at $\alpha$= 0.002 rad.}
\end{figure*}
In this section, we analyze the systematic error caused by the anomaly and put forward some possible solutions to reduce the error. As an example, we analyze the systematic error in the transformation from the measurement with scheme (b) into the measurement with scheme (c). In particular, the anomaly appears when we set a deflection $\alpha$= 0.002 rad on P1 in scheme (c), and the deflection $\beta$ on P2 is the physical quantity to be measured, which can be calculated from the relationship (\ref{weak_value_3}) between the $\beta$ and $\rm [A_w]$. Thereinto, the  $\rm Im$ $\rm [A_w]$ is proportional to the shifts (\ref{delta_p}) of the mean momentum of photons. In the last, the systematic error err($\beta$) of measuring the $\beta$ is shown in Fig. \ref{Fig-system-error}.

\begin{figure*}[htp!]
	\centering
	%\vspace{-0.2cm}
\subfigure
{
	%\begin{minipage}{0.5\linewidth}
	\vspace{-0.2cm}
	\begin{minipage}{6.5cm}
	\centering
	\centerline{\includegraphics[scale=0.55,angle=0]{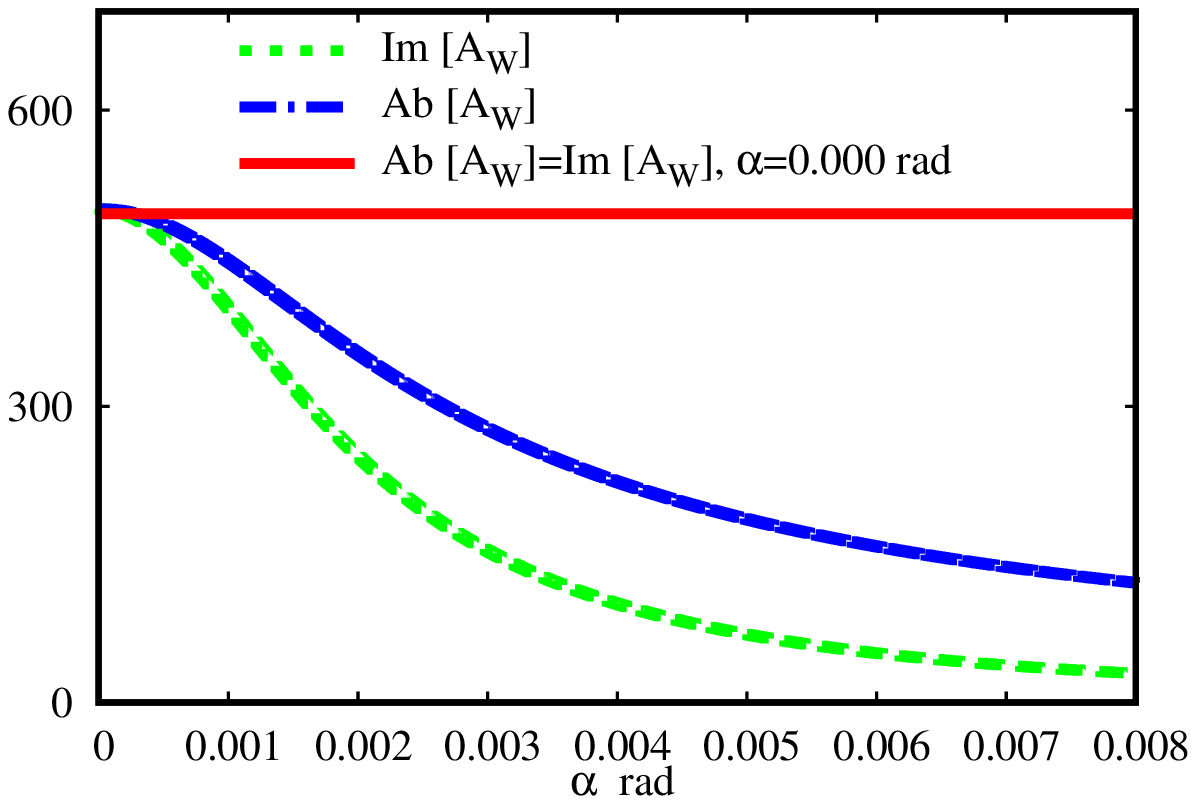}}
	\end{minipage}
}
\vspace{-0.2cm}
\subfigure
{
	%\begin{minipage}{0.5\linewidth}
	\vspace{-0.2cm}
	\begin{minipage}{6.4cm}
	\centering
	\centerline{\includegraphics[scale=0.55,angle=0]{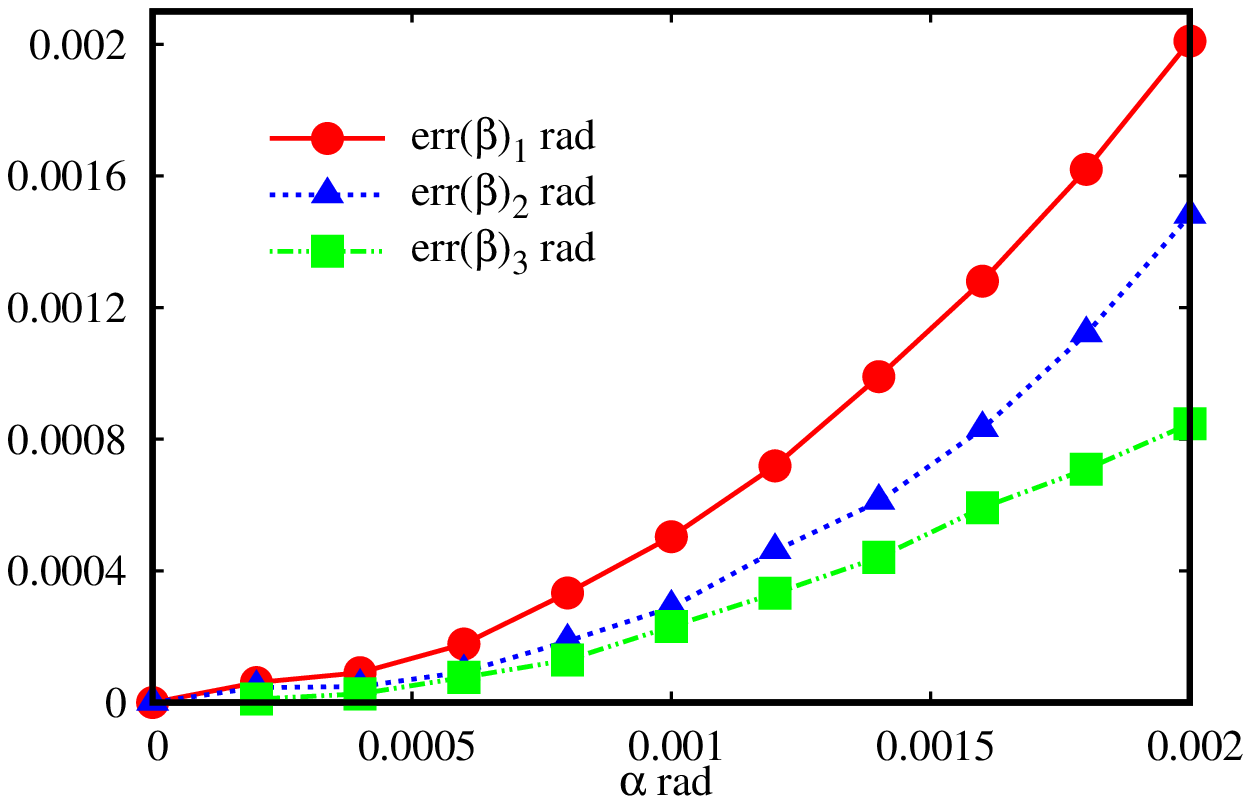}}
	\end{minipage}
}
\vspace*{0mm} \caption{\label{Fig-system-errorde}The weak value (the left panel) and systematic error err($\beta$) (the right panel) dependence of $\alpha$ on the measurement of $\beta$= 0.002 rad .}
\end{figure*}

The numerical simulation of systematic error of $\beta$ is shown in Fig. \ref{Fig-system-error} when measuring the deflection $\beta$ on P2 by considering a deflection $\alpha$= 0.002 rad on P1. Thereinto, the measurement with $\alpha$= 0.000 rad is corresponding to the measurement of scheme (b) as the original design, and the measurement with $\alpha$= 0.002 rad is corresponding to the measurement of scheme (c) with systematic errors caused by the deflection on $\alpha$. By assuming that the true value of $\beta$ is set at 0.002 rad and the value of $\rm Im$ $\rm [A_w]$ is measured at $a_1$. However, because of ignoring the deflection on $\alpha$ in scheme (b), the value of $\beta$ evaluates to 0.0039 rad with the systematic error $err(\beta)_{1}$= 0.0019 rad. On the other hand, the sensitivity which is proportional to the value of $\rm Im$ $\rm [A_w]$ significantly decreases in scheme (c) as $a_1$ is much smaller than $a_4$. 

To sum up, the systematic error caused by the deflection on the optical element in scheme (B) will change the scheme to the scheme (C), as a result of the sensitivity reduction and inaccuracy in the expected measurement with scheme (B). In the next part we will put forward some possible solutions to minish the systematic error:

1). The first plan that came up is to control the parameters of the optical elements precisely even in an unstable environment. Fig. \ref{Fig-system-errorde} shows that the systematic error err($\beta_{1}$) falls off and the difference between $\rm Im$ $\rm [A_w]$($\alpha$= 0.002 rad) and $\rm Im$ $\rm [A_w]$($\alpha$= 0.000 rad) is getting smaller as $\alpha$ decrease. In addition, with the aid of the closed-loop control technique(CLCT) in the automatic control system, CLCT\cite{MILITKY2016335,YAN2021166298} is widely used to accurately control the parameters of the optical elements. Therefore, we can use CLCT to consistently control the polarization of P1 to eliminate the deflection.

2). Note that the difference $a_4$ - $a_2$ between $\rm Ab$ $\rm [A_w]$($\alpha$= 0.002 rad) and $\rm Im$ $\rm [A_w]$($\alpha$= 0.000 rad) is smaller than the difference $a_4$ - $a_1$ between $\rm Im$ $\rm [A_w]$($\alpha$= 0.002 rad) and $\rm Im$ $\rm [A_w]$($\alpha$= 0.000 rad), while the systematic error err($\beta_{2}$) estimated by $\rm Ab$ $\rm [A_w]$ is smaller than the systematic error err($\beta_{1}$) estimated by $\rm Im$ $\rm [A_w]$. Similarly, following the method to estimate err($\beta_{1}$) and err($\beta_{2}$) in the left panel of Fig. \ref{Fig-system-error}, the err($\beta_{1}$) and err($\beta_{2}$) dependence of $\alpha$ are shown in the right panel of Fig. \ref{Fig-system-errorde}. Therefore, using $\rm Ab$ $\rm [A_w]$ to calculate the quantity $\beta$ can get a smaller systematic error err($\beta$) compared to the traditional method with $\rm Im$ $\rm [A_w]$. In addition, it is feasible and convenient to measure both the shifts of the position and the momentum of the photon in practical measurements, and this method can effectively reduce the systematic error at certain deflection on the optical element.

3). From Fig. \ref{Fig-system-error} and Fig. \ref{Fig-system-errorde}(right panel), we can conclude that the systematic error err($\beta_{3}$) estimated by the probability $\mathcal{P}$ of obtaining a post-selected state is much smaller than err($\beta_{1}$) obtained by $\rm Im$ $\rm [A_w]$. Note that Li etc.\cite{PhysRevA.102.023701} have proposed a scheme called the weak measurement with two pointers to measure time delays with the light frequency and the post-selected probability $\mathcal{P}$ in dynamic range with high precision. Thus, the similar scheme with measurement of the probability $\mathcal{P}$ consulting their work\cite{PhysRevA.102.023701} can availably reduce the systematic error.

In addition, one thing to note in weak measurements is that estimating the $\rm Im$ $\rm [A_w]$ or $\rm Re$ $\rm [A_w]$ in the scheme with a plural weak value is improper, due to the abnormal non-monotone behaviour of $\rm Im$ $\rm [A_w]$ or $\rm Re$ $\rm [A_w]$. Note that the sensitivity in WVA is proportional to $\rm Im$ $\rm [A_w]$ or $\rm Re$ $\rm [A_w]$, the abnormal non-monotone behaviour will evaluate two physical quantities. As shown in Fig. \ref{Fig-system-error}(left panel), if the $\rm Im$ $\rm [A_w]$= $a_5$ is calculated from the shift of the momentum of the photons, the value of $\beta$ to be measured will get two values $\beta_{1}$ and $\beta_{2}$. In conclusion, the weak measurement should be realized with a pure real $\rm [A_w]$ or pure imaginary $\rm [A_w]$, and the above three methods are effective and necessary to reduce systematic error and enhance the sensitivity when the weak measurement changed to measurement with the plural $\rm [A_w]$.

\section{Conclusion}
We have found the systematic error and the anomaly in the sensitivity and the probability of post-selection in weak measurement raised from a plural weak value. When we realize a scheme of weak measurement with a pure real $\rm [A_w]$ or a pure imaginal $\rm [A_w]$, the anomaly caused by the deflection on polarizer will change the scheme into the measurement with a plural weak value, and the anomaly will significantly reduce the sensitivity and generate a systematic error of the measurement compared with the original scheme. 

In the end, we provide three feasible methods to reduce the systematic error and decrease the sensitivity: minimizing the parameter error of optical elements as much as possible by CLCT, using the modulus of $\rm [A_w]$ or the probability of obtaining a post-selected state to estimate the physical quantity to be measured rather than using the imaginal part of $\rm [A_w]$. In addition, the joint measurement with the modulus of $\rm [A_w]$ and the probability of obtaining a post-selected state will be our next step to reduce these negative effects when the weak measurement inevitably changes into measurement with a plural weak value.

\section{Backmatter}

\begin{backmatter}

\bmsection{Acknowledgments}
This study is financially supported by MOST Special Fund from the State Key Laboratory of Geological Processes and Mineral Resources, China University of Geosciences No. MSFGPMR01-4 and the Fundamental Research Funds for National Universities, China University of Geosciences(Wuhan) (Grant No. G1323519204).

\bmsection{Disclosures}

\noindent  The authors declare no conflicts of interest.

\bmsection{Data Availability Statement}
 Data underlying the results presented in this paper are not publicly available at this time but may be obtained from the authors upon reasonable request.

\end{backmatter}

\bibliography{sample}
\end{document}